\documentclass{osa-article}
\usepackage{subcaption}
\usepackage{wrapfig}
\usepackage{floatflt}
\usepackage{siunitx}
\usepackage{graphicx}
\usepackage{float}

\journal{oe}

\articletype{Research Article}

\begin{document}

\title{Thermal tuning of a fiber-integrated Fabry-Pérot cavity}

\author{Clemens Singer,\authormark{1,3} Alexander Goetz,\authormark{1,3} Adarsh S. Prasad,\authormark{1} Martin Becker,\authormark{2} Manfred Rothhardt,\authormark{2}  and Sarah M. Skoff\authormark{1,4}}

\address{\authormark{1}Vienna Center for Quantum Science and Technology, Atominstitut, Technische Universität Wien, Stadionallee 2, 1020 Vienna, Austria\\
\authormark{2}Leibniz Institute of Photonic Technology, Albert-Einstein-Strasse 9, 07745 Jena, Germany\\}

\address{\authormark{3} Co-first authors with equal contribution}
\email{\authormark{4}sarah.skoff@tuwien.ac.at}

\section{Abstract}

Here, we present the thermal tuning capability of an alignment-free fiber-integrated Fabry-Perot cavity. The two mirrors are made of fiber Bragg gratings that can be individually temperature stabilized and tuned. We show the temperature tuning of the resonance wavelength of the cavity without any degradation of the finesse and the tuning of the individual stop bands of the fiber Bragg gratings. This not only permits for the cavity's finesse to be optimized post-fabrication but also makes this cavity applicable as a narrowband filter with a FWHM spectral width of $0.07\pm0.02$ pm and a suppression of more than -15 dB that can be wavelength tuned. Further, in the field of quantum optics, where strong light-matter interactions are desirable, quantum emitters can be coupled to such a cavity and the cavity effect can be reversibly omitted and re-established. This is particularly useful when working with solid-state quantum emitters where such a reference measurement is often not possible once an emitter has been permanently deposited inside a cavity.

\section{Introduction}

Fiber Bragg gratings are used for a variety of technological applications due to their small size, their compatibility with optical fibers which naturally comes with an ease in signal transmission and their ability for multiplexing. They have been used as temperature \cite{tao_sensor_2017}, refractive index \cite{ferreira_fabry-perot_2017,zhou_refractometer_2011,gouveia_fabryperot_2012} , strain \cite{markowski_linearly_2017} magnetic fields \cite{zhao_fiber_2014} and pressure sensors \cite{Fiber_anemometer,Othonos} and for measurements of windspeed, liquid levels, liquid density and specific gravity \cite{lai_application_2012}. By using two gratings to form a Fabry-Perot cavity they have also shown the potential to function as narrowband filters\cite{zhang_silicon-based_2015,zhou_optical_2015}. Their suitability for forming an alignment-free Fabry-Perot type cavity \cite{wei_subkilohertz_2016} has also opened up their application in the field of quantum optics, where the fiber connecting two such fiber Bragg gratings can be readily tapered to guide the light between the gratings as an evanescent wave \cite{hutner_nanofiber-based_2020,wuttke_nanofiber_2012,schell_highly_2015,kato_strong_2015}. This allows for efficient coupling of quantum emitters to the guided light field of such nanofiber-based cavities. For all these applications, a tunability of the fiber Bragg gratings is desirable especially when it is required to match the resonance of a specific quantum emitter or transmit or reflect light in a narrow wavelength range \cite{hedger_high_2020}. For that purpose, complex designs exist to electrically tune a fiber Bragg grating cavity \cite{zhang_silicon-based_2015}. Here, we demonstrate the thermal tunability of a fiber Bragg grating Fabry-Perot cavity, where each fiber Bragg grating can be individually temperature stabilized and tuned. This simple design not only enables to find the perfect overlap of the stop bands of the individual fiber Bragg gratings and thus highest finesse of the cavity post-fabrication but also allows for tuning the stop band of the two fiber Bragg gratings far enough such that the cavity effect can be reversibly suspended and re-established. This is particularly useful for coupling solid-state quantum emitters to such cavities as it allows for reference measurements with and without a cavity structure which is otherwise not possible once an emitter is permanently deposited on a cavity. In addition it finds its applicability in quantum optics to separate probe and control fields that are often narrowband and may be only separated by a few GHz \cite{hedger_high_2020}. This infiber cavity therefore conveniently allows the suppression of one beam compared to the other by more than -15dB which can be easily increased by concatenating such infiber cavities without the need for alignment.

\section{Experimental Setup}

The Fabry-Perot cavity is written into a commercial single mode fiber (SM600) by a pulsed 248nm KrF laser that generates the two fiber Bragg gratings at a distance of 25 mm from each other \cite{Lindner:09}. To be able to stabilize and temperature tune the two gratings individually, two PID controllers (Arroyo Instruments) are used in conjunction with two Peltier elements and two thermistors that cool or heat and monitor the temperature, respectively. 
\begin{figure}[H]
\centering
\includegraphics[ height=5cm]{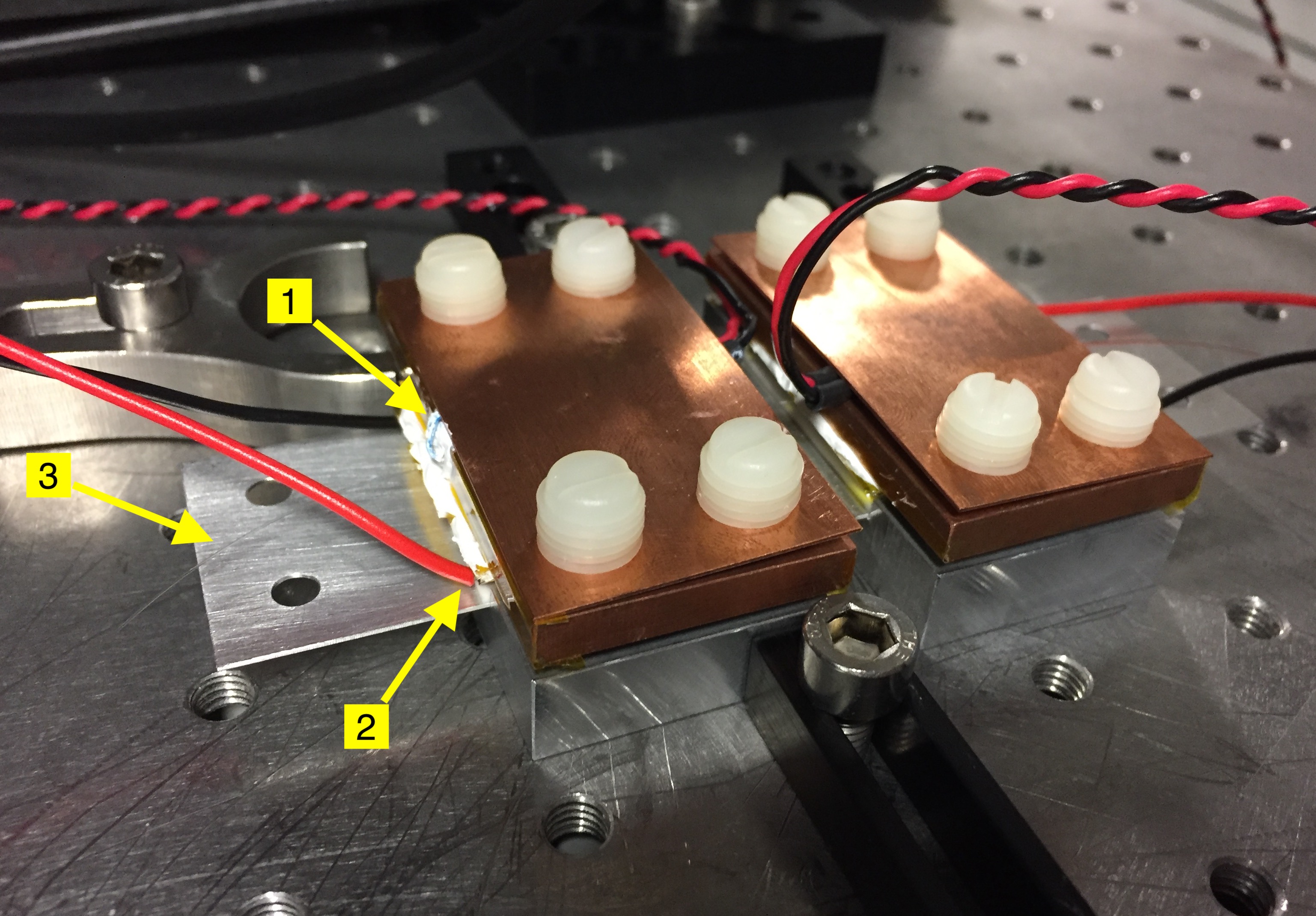} 
\caption{Fiber mount with (1) the thermistor in the hole of the top plate, (2) the Peltier element on the bottom and (3) the fiber itself.}
\label{pic:fiber-mount}
\end{figure}
There are other heating techniques that can in principle be used, such as internal laser heating \cite{Fiber_anemometer,zhou_optical_2015}  or resistive heating via a thin metal coating around the fiber\cite{Fiber_chirp} but these methods do not allow for cooling of the fiber grating. In addition, the fiber Bragg gratings can also be tuned by strain but all these methods require more sophisticated set-ups \cite{liu_free-spectral_2009}. In our case, the cavity holder (Fig.\ref{pic:fiber-mount}) consists of a U-shaped ground plate, made from aluminium, and two copper plates on top, that keep the Peltier elements and the fiber in place. The Peltier elements are fixed in the mount by grooves, just deep enough for the elements to stay in place but not too deep, so that the top and bottom parts of the Peltier elements are not thermally coupled via the aluminium holder. The fiber is placed between the Peltier elements and copper plates. Copper in conjunction with thermal silicon grease was used to ensure an even distribution of heat across the whole length of the fiber beneath. The temperature sensors are then clamped on the top copper plates with two additional thin copper plates. This cavity is characterized by a widely tunable laser (New Focus TLB 6700) where the transmission is monitored by two photodiodes or single photon counting modules. Before the light enters the cavity it is passing through three polarization paddles, which can be tilted to adjust the polarization of the light. This is necessary to end up with only one resonant mode inside the cavity \cite{wuttke_nanofiber_2012}. Before the beam enters the cavity some light is picked off to monitor the wavelength via a wavemeter which continuously monitors the laser frequency with a relative precision better than $3 \times 10^{-7}$.

\section{Results}
\subsection{Individually tuning the stop bands of the two fiber Bragg gratings}

\begin{figure}[H]
\centering
\includegraphics[width=0.8\linewidth]{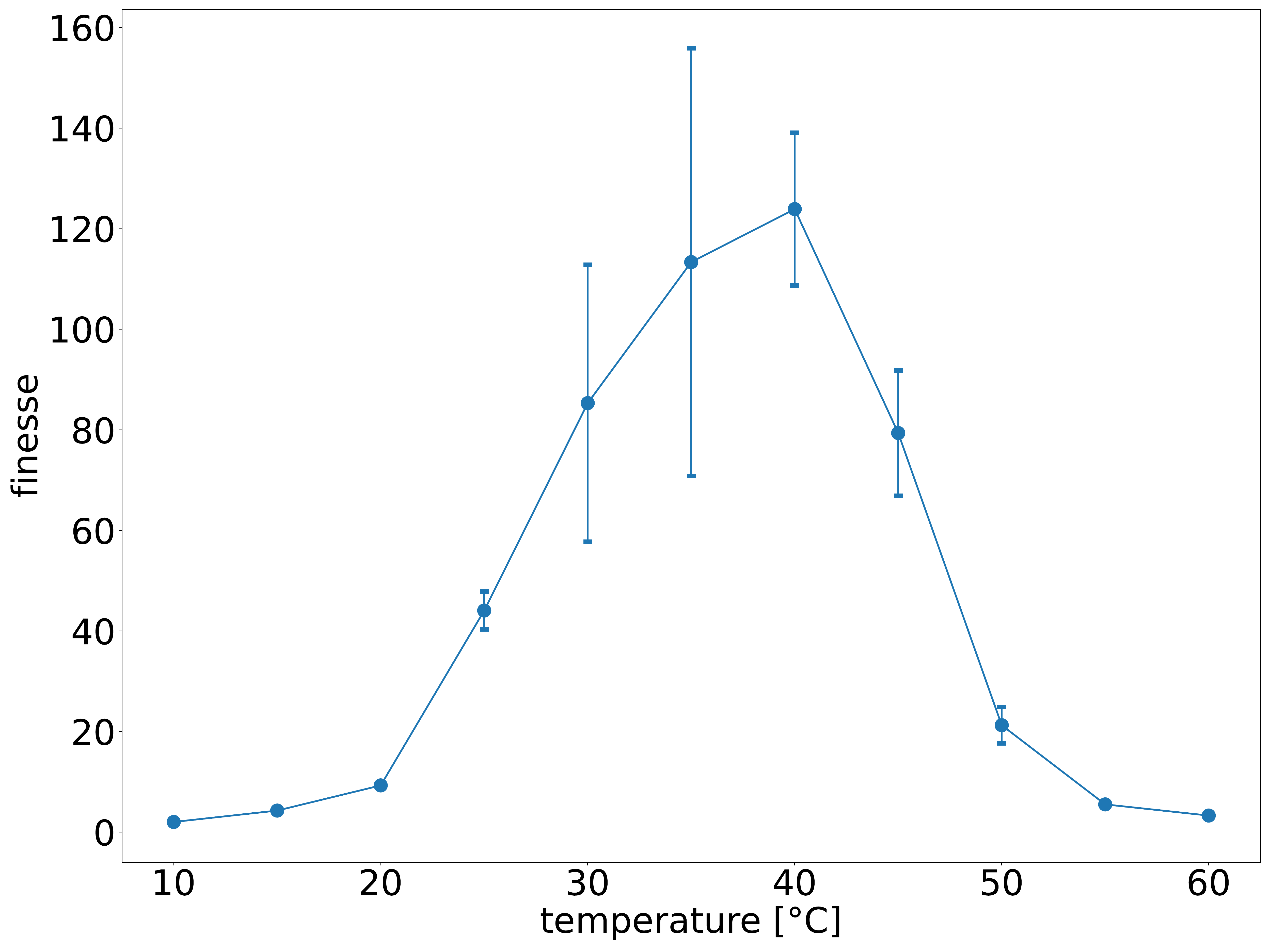} 
\caption{Maximum finesse values (a) in transmission when grating one is heated from 10 to 60°C. With grating two, constantly held at 22°C.}
\label{fig:maxfintrans}
\end{figure}

Although the Bragg gratings are manufactured for a particular Bragg wavelength $\lambda_\text{B}$ given by $\lambda_\text{B} = 2  n_{\text{eff}} \, \Lambda$, where $\Lambda$ is the grating period and $n_{\text{eff}}$ the effective refractive index, there might still be a slight mismatch between two gratings after manufacturing that results in a lower finesse than could be achieved. To maximise the overlap between the two gratings, one is held at a constant temperature of 22°C as the other grating is heated from 10°C to 60°C.  Once there is an overlap between the two stopbands and a cavity effect can be observed, the finesse $F$ is obtained by fitting an Airy function 
\begin{equation}
    T(\lambda)=const+\frac{T_0}{1+\frac{2F^2*sin^2(\frac{\pi(\lambda-\lambda_0)}{\lambda_{FSR}})}{\pi}},
    \label{eq:Airy}
\end{equation}
to two spectral lines at a time. The highest finesse at a particular wavelength then indicates the wavelength of best overlap of the two stop bands at a particular temperature of the second grating \cite{hutner_nanofiber-based_2020}.The finesse  is plotted as a function of grating temperature in Fig. \ref{fig:maxfintrans}. Each datapoint in Figure \ref{fig:maxfintrans} corresponds to the mean of three measurements and the errorbars represent its standard deviation. It can be seen, that the highest finesse value and thus the best overlap of the two stopbands is reached, when grating one is heated to 37°C, where the finesse reaches $124 \pm 15$.

\begin{figure}[H]
\begin{subfigure}{0.5\textwidth}
    \includegraphics[width=1\linewidth]{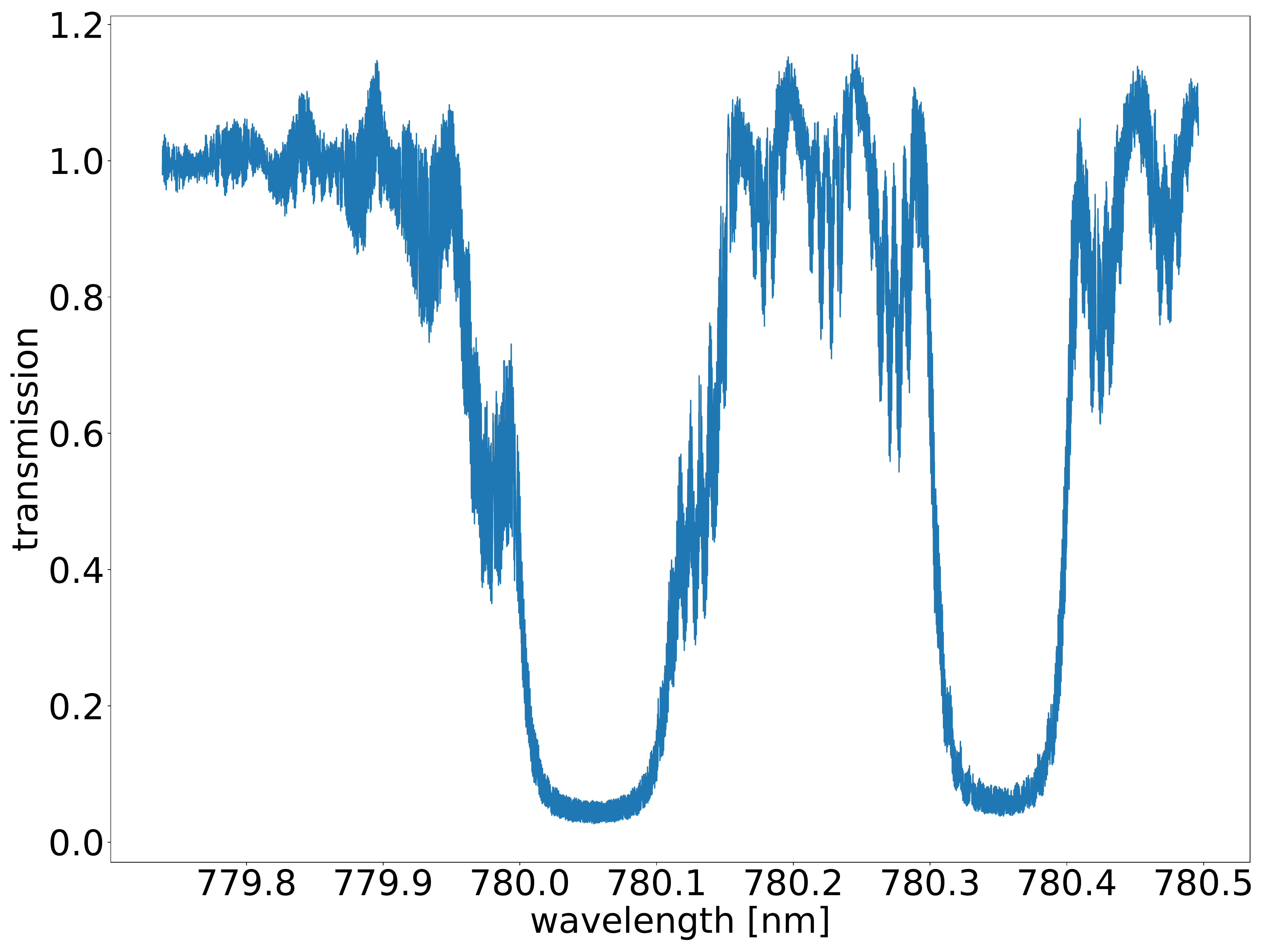}
    \caption{}
    \label{fig:worstoverlap}
\end{subfigure}
\begin{subfigure}{0.5\textwidth}
    \includegraphics[width=1\linewidth]{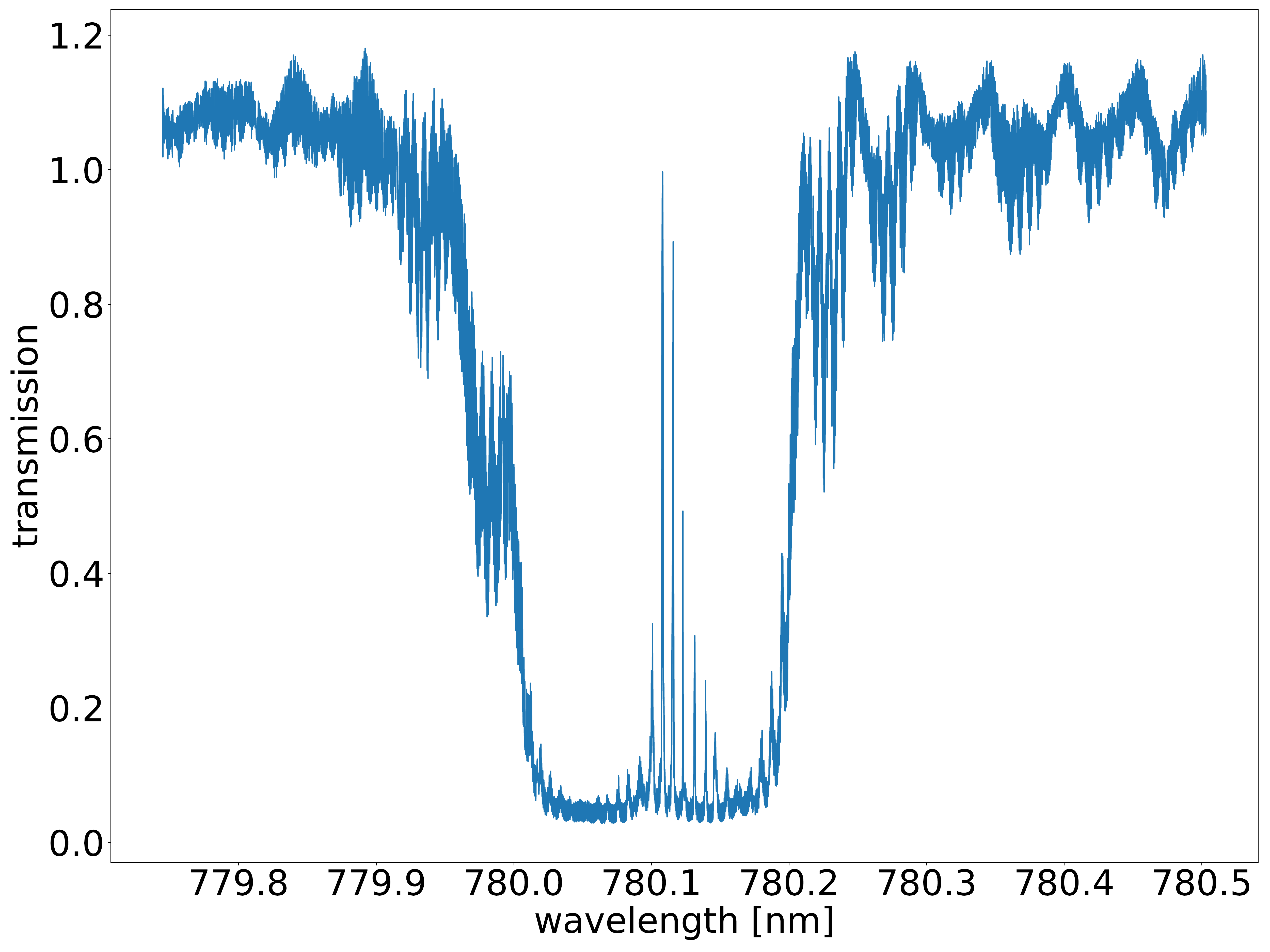}
    \caption{}
    \label{fig:midoverlap}
\end{subfigure}
\begin{subfigure}{0.5\textwidth}
    \includegraphics[width=1\linewidth]{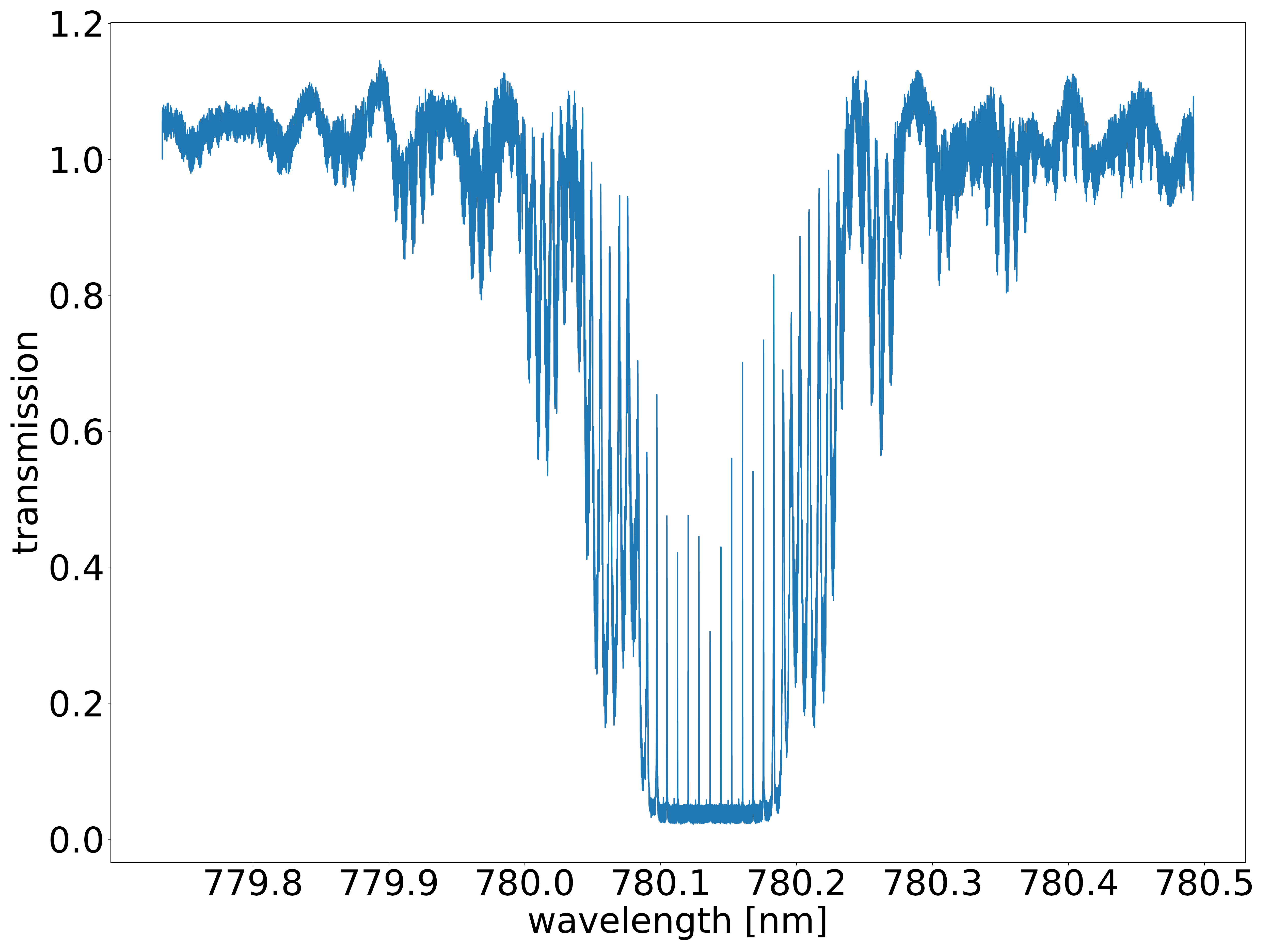}
    \caption{}
    \label{fig:bestoverlap}
\end{subfigure}
\begin{subfigure}{0.5\textwidth}
    \includegraphics[width=1\linewidth]{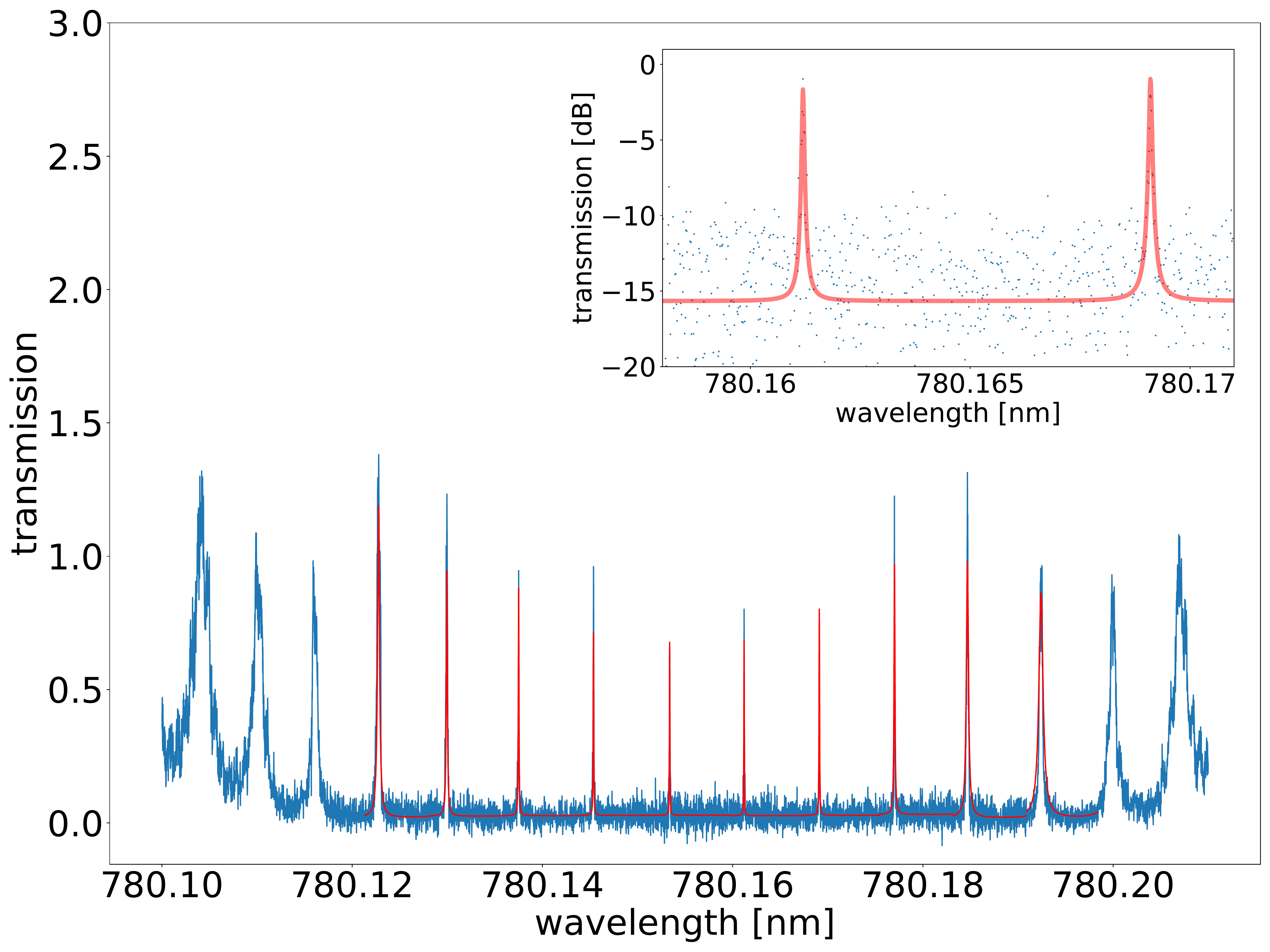} 
    \caption{}
    \label{fig:dBplot}
\end{subfigure}
\caption{Transmission spectra with grating two at 60°C (a) showing two separate stop bands, with grating one and two at room temperature (b) showing the finesse that is obtained without any temperature tuning and grating one at 37°C (c) showing best overlap. d) zoomed in region of transmission peaks at best overlap with an inset that shows the transmission plotted with a db scale to show functionality as a narrowband filter}
\label{fig:overlapspectra}
\end{figure}

 By heating one grating and cooling the other we are able to achieve an even larger detuning of the two grating stop bands and in Fig \ref{fig:overlapspectra}, three cases are depicted. In (Fig. \ref{fig:overlapspectra} a). the detuning is large enough that the two individual stopbands of the Bragg gratings can be seen and  the cavity effect is eliminated. Fig. \ref{fig:overlapspectra} b) show the overlap with both gratings at room temperature and  Fig. \ref{fig:overlapspectra} c) shows the transmission spectrum of the cavity with an optimized overlap of the individual stopbands. In Fig. \ref{fig:overlapspectra} d), a zoom in of the transmission peaks at best overlap as in c) is displayed including a fit to eqn. \ref{eq:Airy} and an inset which shows the transmission in dB to infer the possible suppression of unwanted light by this type of filter.

\subsection{Tuning the center frequency of the cavity}
Whether such a cavity is used as a narrowband filter or to  enhance the light-matter interaction between a light field and a quantum emitter, it may be necessary to tune the center frequency of the cavity. The Bragg wavelength $\lambda_\text{B}$ of a Fiber-Bragg-Grating increases when the grating is heated. This effect originates from two sources. Firstly, thermal expansion of the fiber itself causes an increase of the distance between each grating and secondly the effective refractive index of the material $n$ changes with temperature.
Both these effects on the Bragg wavelength are described by the following equation,
\begin{equation}
    \Delta\lambda_\text{B}=\lambda_\text{B}(\alpha_T+\alpha_n)\Delta T,
\end{equation}
where \(\alpha_T\) is the thermal expansion coefficient and \(\alpha_n\) the thermo optic coefficient accounting for the change in effective refractive index. By tracking the position of narrow features in the reflection spectra, it is possible to obtain the wavelength shift of the two gratings as a function of temperature. It has been previously demonstrated that the temperature-dependent wavelength shift for a temperature range between 10-60 degrees is linear with temperature \cite{FBG_shift_low_temperatures,Othonos,hill_fiber_1997,Thermooptic_coeff} and hence the resulting datapoints were fitted with a linear function\cite{FBG_shift_low_temperatures} that showed a shift of $6.3 \pm 0.3$ pm/K for grating one and $5.0 \pm 0.2$ pm/K for grating two. These values are in agreement with the temperature-dependent wavelength shift of the Bragg grating in silica \cite{Othonos,Thermal_expansion_coeff,Thermooptic_coeff,FBG_shift_low_temperatures}. In principle this shift can be enhanced further by bonding the fiber to a substrate with a large thermal expansion coefficient such as copper, in which case we would expect the shift to increase to 13.41 pm/K \cite{Thermal_expansion_coeff_copper}. In our current set-up by just fixing the fiber losely with some Kapton tape on the Peltier element, and then clamping it with the copper plate, as expected we see no influence of the surrounding material on the wavelength shift.

\begin{figure}[H]
\begin{subfigure}{0.5\textwidth}
\includegraphics[width=0.9\linewidth]{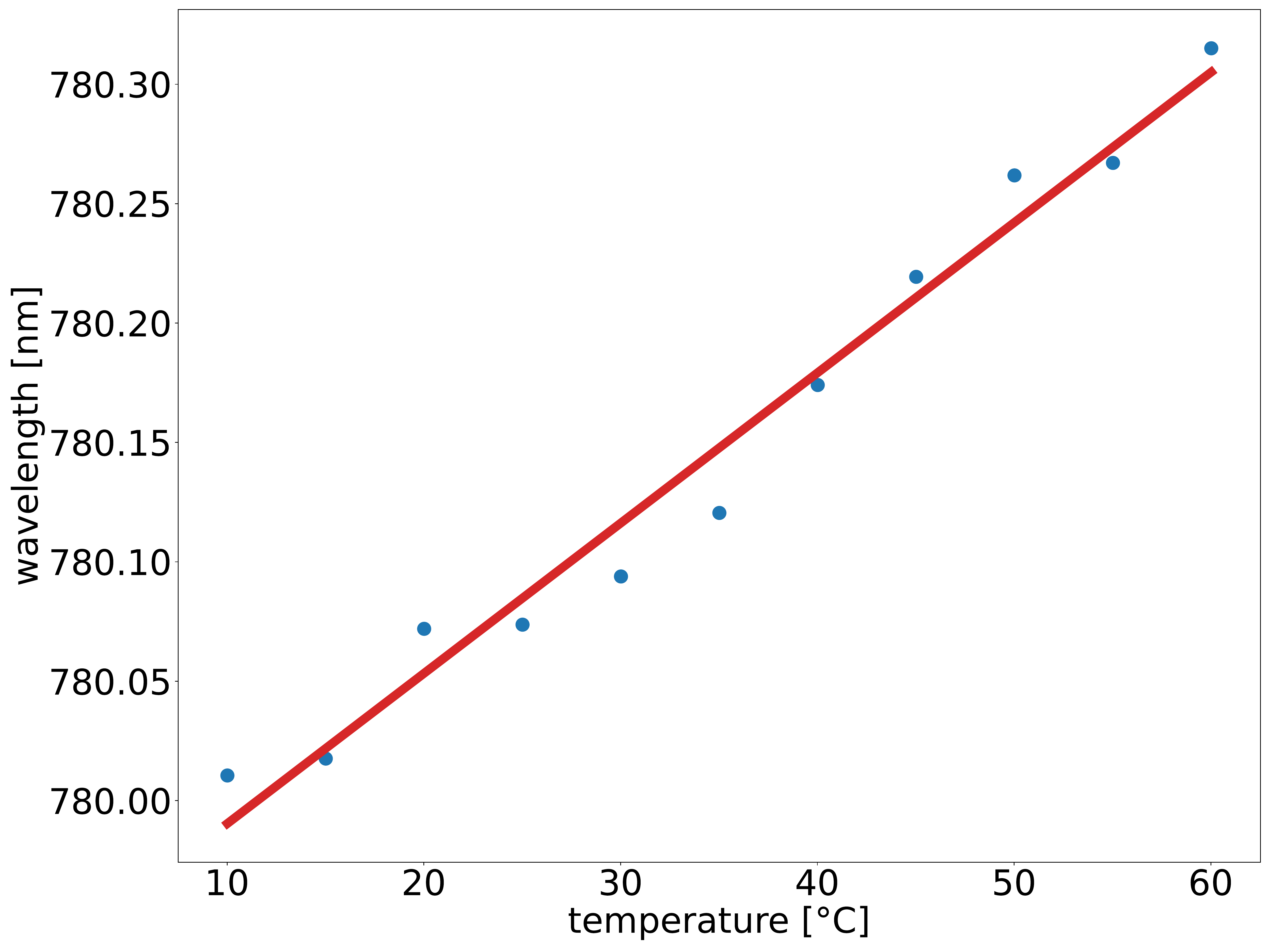} 
\caption{Shift of grating one}
\label{fig:wavelengthshiftgr1}
\end{subfigure}
\begin{subfigure}{0.5\textwidth}
\includegraphics[width=0.9\linewidth]{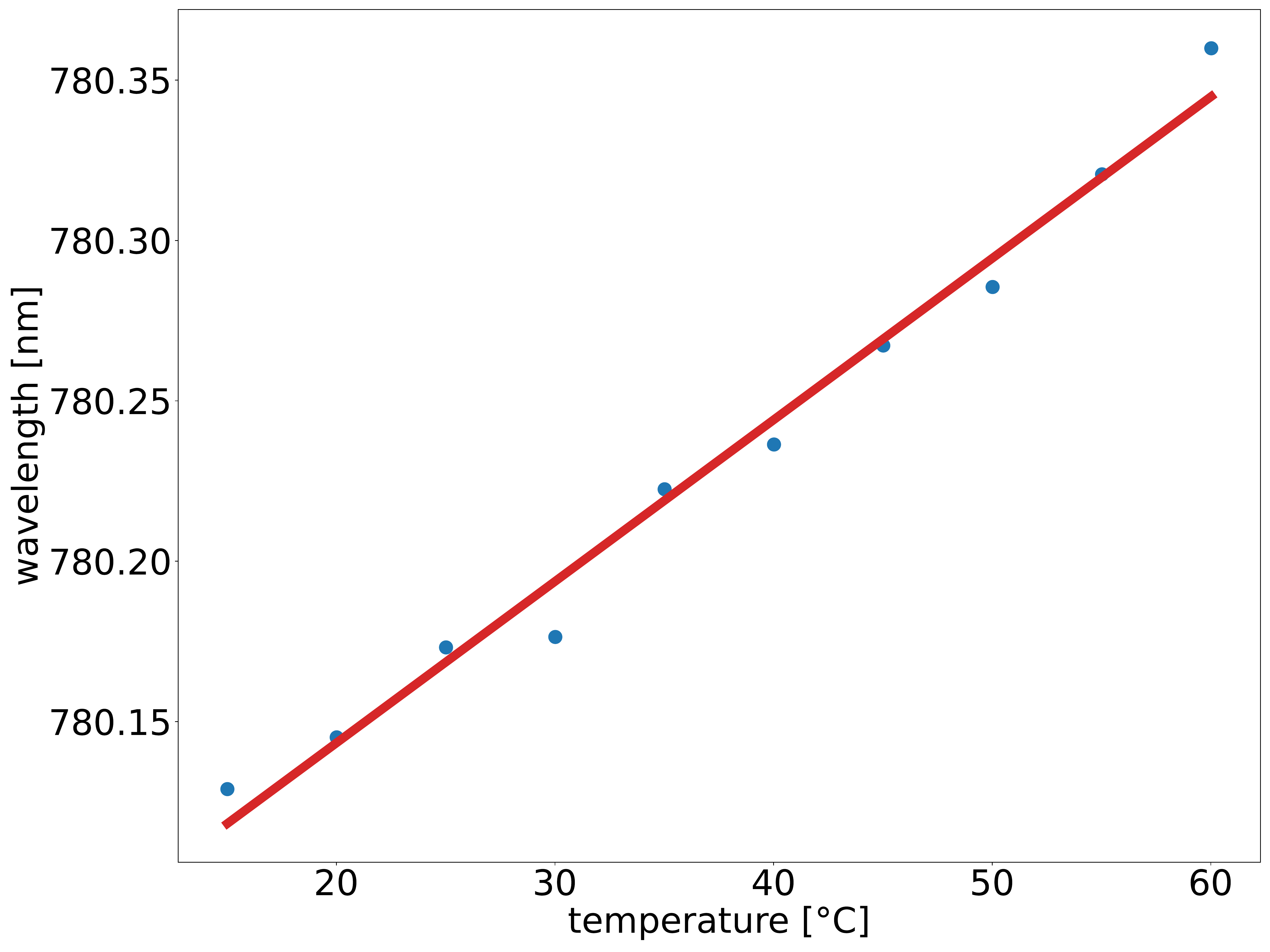}
\caption{Shift of grating two}
\label{fig:wavelengthshiftgr2}
\end{subfigure}
\caption{Wavelength shift when (a) grating one and (b) grating two is heated. Fitted function in red.}
\label{fig:wavelengthshift}
\end{figure}

\begin{figure}[H]
\centering
\includegraphics[width=0.9\linewidth]{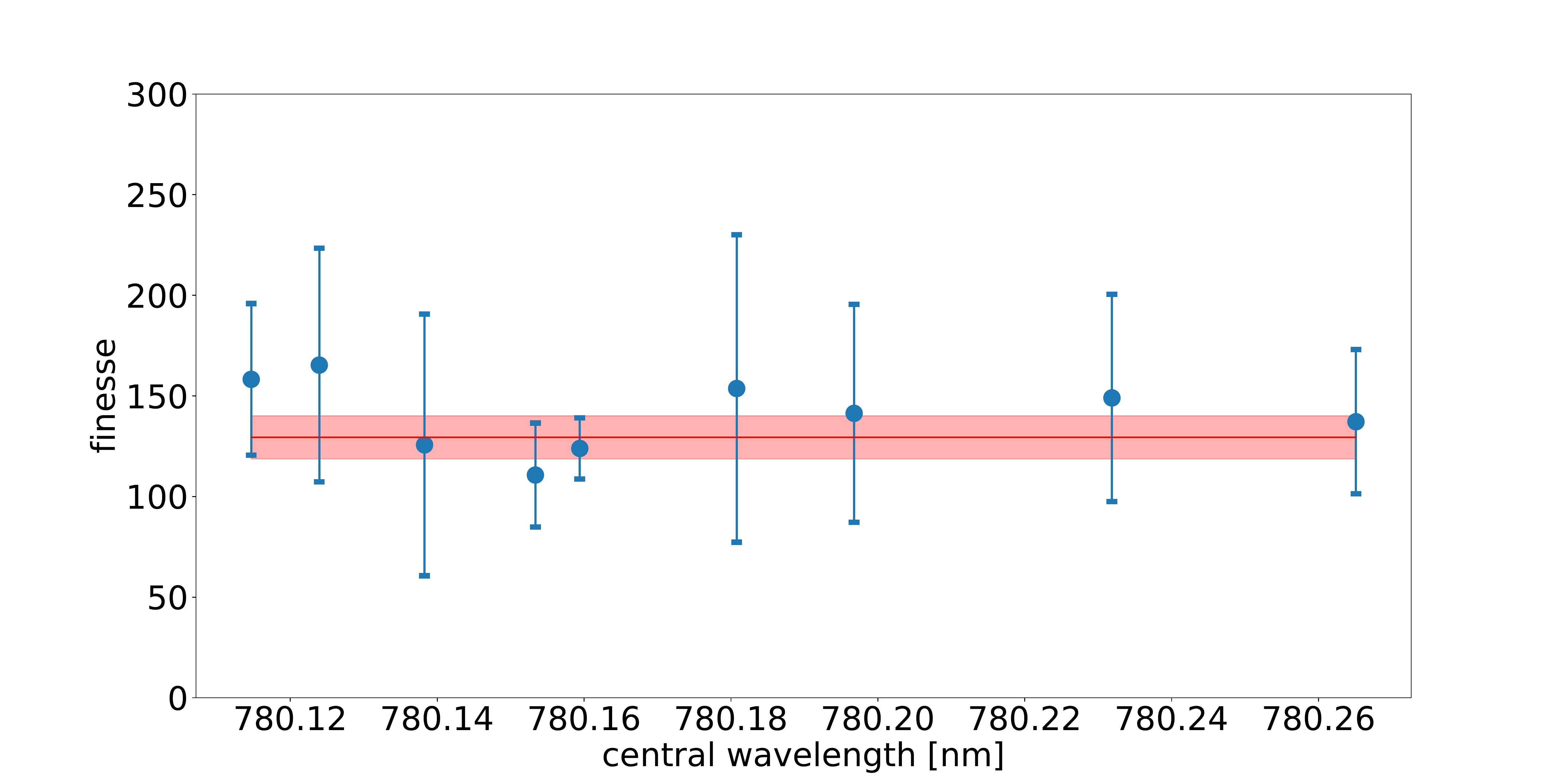} 
\label{fig:mainmaxfintrans}
\caption{Maximum finesse values as the center frequency of the cavity is temperature-tuned. The datapoints are the mean value of three measurements and the errorbars are the corresponding standard deviation}
\label{fig:mainmaxfin}
\end{figure}

The measurements of the wavelength shift (Fig. \ref{fig:wavelengthshift}), together with the known point of maximum overlap (Fig. \ref{fig:maxfintrans}) and hence largest finesse, makes it possible to temperature control both gratings in such a way that the overlap is kept at the optimum with an average finesse of  $129 \pm 11$ across a tuning range of about 160 pm as depicted in Fig. \ref{fig:mainmaxfin}.

\section{Conclusion}

In summary, we have demonstrated a fiber-integrated cavity that can be temperature tuned over hundreds of pm. This tuning allows for finding the best overlap between the two fiber Bragg gratings to optimise the finesse even after the gratings have been fabricated. The temperature control together with the finite width of the stop bands of the individual fiber Bragg gratings that function as the mirrors of this Fabry-Perot type cavity has further allowed us to reversibly and controllably omit and re-establish the cavity effect. This can be especially important when coupling solid-state quantum emitters to such cavities as can be readily done by placing an emitter on a tapered fiber section between the gratings \cite{hutner_nanofiber-based_2020}, thereby creating a high-Q alignment-free cavity. This tunability then allows for convenient reference measurements with and without the cavity, something which is otherwise often not possible with permanently deposited solid-state quantum emitters. The tuning range of the cavity's center frequency without any degradation of the finesse allows coupling to solid-state emitters such as colour centers in diamond \cite{Narrow-linewidth_diamond_emitters}, quantum emitters in 2D materials\cite{tran_resonant_2018} or single molecules in solids\cite{pazzagli_self-assembled_2018,skoff_optical-nanofiber-based_2018} that have an inherent inhomogeneous broading due to their nanoenvironment. In the broader context of photonics, the demonstrated tunability of the individual stop bands further allows this cavity to function as a wavelength tunable narrowband filter that is alignment-free and can be easily incorporated in any experiment. Our results therefore promise a wide applicability of these fiber-integrated Bragg grating cavities not only in quantum optics but also for new photonic technologies.

\section{Funding Information}
Austrian Academy of Sciences (Quantum Light No. 1847108);  European
Commission (project ErBeStA No. 800942)

\bibliography{FBG_references,cavityBib,allRef}






\end{document}